\documentstyle[aps,prl,floats]{revtex}
\input{epsf}
\def \ind#1{{\mbox{\footnotesize {#1}}}}
\def \ind#1{{\mbox{\scriptsize {#1}}}}
\input{psfig}

\begin{document}
\draft

\twocolumn[\hsize\textwidth\columnwidth\hsize\csname@twocolumnfalse\endcsname
\title{Dynamical Heterogeneities Below the Glass Transition}
\author{K. Vollmayr-Lee$^{1,2}$, W. Kob$^3$, K. Binder$^2$
             and  A. Zippelius$^4$}
\address{$^1$Department of Physics, Bucknell University,  
      Lewisburg, Pennsylvania 17837, USA\\
  $^2$Institute of Physics, Johannes-Gutenberg-University Mainz,
      Staudinger Weg 7, 55099 Mainz, Germany\\
  $^3$Laboratoire des Verres, cc69, Universit\'e  Montpellier II
   34095 Montpellier Cedex 05, France\\
  $^4$Institute of Theoretical Physics, Georg-August-University G\"ottingen,
      Bunsenstr. 9, 37073 G\"ottingen, Germany}
\date{\today }
\maketitle

\begin{abstract}
We present molecular dynamics simulations 
of a binary Lennard-Jones mixture at temperatures below the kinetic 
glass transition. The ``mobility'' of a particle is
characterized by the amplitude of its fluctuation around its
average position. The 5\% particles with the largest/smallest mean amplitude
are thus defined as the relatively most mobile/immobile particles.
We investigate for these 5\% particles their spatial distribution and
find them to be distributed very heterogeneously in that mobile as well
as immobile particles form clusters. The reason for this dynamic
heterogeneity is traced back to the fact that
mobile/immobile particles are surrounded by fewer/more neighbors which
form an effectively wider/narrower cage. The dependence of our
results on the length of the simulation run indicates that individual 
particles have a characteristic mobility time scale, which can be
approximated via the non-Gaussian parameter.
\end{abstract}

]

\section{Introduction}

\label{sec:intro} 
Although glasses have already been studied for a long time, their
complete understanding is still an open problem due to the complex behavior
of their static and dynamic quantities \cite{ref:glassintro}. We focus here
on their dynamics which has been found to relax non-exponentially in the
supercooled liquid and shows strong history
dependence below the glass transition.\\
The question arises whether this behavior is due to spatially 
homogeneous non-exponential dynamics or spatially heterogeneous
dynamics (for review articles see 
\cite{ref:ediger,ref:sillescu,ref:bohmer}).
Since a glass is an amorphous solid and therefore not all atoms 
are structurally equivalent, one should expect that also their 
dynamics differs, i.e.\ that the system has a heterogeneous dynamics. 
Recently the answer to the question of dynamic heterogeneity has been
addressed both by means of experiments 
\cite{ref:nonexp,ref:kegel,ref:eweeks} 
and computer simulations of two-dimensional  
\cite{ref:comp_2d,ref:alpha2_1,ref:gould} and three-dimensional systems
\cite{ref:yamamoto,ref:doliwa+heuer,ref:laird,ref:schroeder,ref:nistwork,ref:oligschleger}.
Here we study a binary Lennard-Jones mixture in three dimensions
which has been investigated before  extensively
\cite{ref:nistwork} and which shows clear dynamic heterogeneity
{\em above} the glass transition. Similar dynamics has been found
experimentally with confocal microscopy of a colloidal suspension 
in the supercooled fluid and in the glass \cite{ref:kegel,ref:eweeks}. 
Whereas most simulations have been done at relatively high temperatures
{\em above} the calorimetric glass transition, and the 
experiments of atomic systems
were performed {\em near} the glass transition, we simulate, in this
paper, {\em below} the glass transition (which has so far only been done 
experimentally by Weeks {\it et al.\ }\cite{ref:eweeks} and 
in simulations by Oligschleger {\it et al.\ }\cite{ref:oligschleger}).
We find here dynamic heterogeneity via simulations of the same binary
Lennard-Jones system as \cite{ref:nistwork} but {\em below} the glass
transition.
In contrast to the previous simulations we have the picture of a solid 
in mind, instead of coming from the liquid. 
We use the ``localization length'' of the work of reference
\cite{ref:goldbart+zippelius} to define the mobility of a
particle as the mean fluctuation around its average position.
To address the question of what allows or inhibits a particle to be mobile,
we study the surrounding of these particles. Using different lengths
of simulation runs we also learn about the time scale over which mobile and
immobile particles sustain their character. 

We review in section \ref{sec:simulation} the model used and give
details of the simulation. In section \ref{sec:mobile_immobile} we
present the mean square displacement and the mean fluctuations of 
a particle around its average position and define what we mean by 
mobile and immobile particles. We then study their spatial distribution
(sec. \ref{sec:spatialdist}), surrounding (sec. \ref{sec:surrounding}) and
time scale (sec. \ref{sec:timescale}) and conclude with section
\ref{sec:summary}.

\section{Simulation Details}  \label{sec:simulation}

We study a binary Lennard-Jones (LJ) mixture of 800 A and 200
B particles. Both A and B particles have the same mass. The
interaction between two particles $\alpha$ and $\beta$ 
($\alpha, \beta \in \{$A,B$\}$) is 
\begin{equation}  
V_{\alpha \beta}(r) = 4 \, \epsilon_{\alpha \beta} \,
\left ( \left ( \frac{\sigma_{\alpha \beta}}{r} \right )^{12}
      - \left ( \frac{\sigma_{\alpha \beta}}{r} \right )^{6}
\right ),
\end{equation}
where $\epsilon_\ind{AA}=1.0 , \epsilon_\ind{AB}=1.5 , \epsilon_\ind{BB}=0.5 ,
\sigma_\ind{AA}=1.0 , \sigma_\ind{AB}=0.8$ and
$\sigma_\ind{BB}=0.88$. We truncate and shift the potential at 
$r=2.5 \sigma_{\alpha \beta}$. 
From previous investigations \cite{ref:kob_andersen} it is known
that this system is not prone to crystallization and demixing. 
In the following we
will use reduced units where the unit of length is $\sigma_\ind{AA}$,
the unit of energy is $\epsilon_\ind{AA}$ and the unit of time is 
$\sqrt{m \sigma_\ind{AA}^2/(48 \epsilon_\ind{AA})}$. 

We carry out molecular dynamics (MD) simulations using the velocity
Verlet algorithm with a time step of 0.02. The volume is kept constant
at $V=9.4^3=831$ and we use periodic boundary conditions. We are
interested in the dynamics of the system below the glass transition. 
Since recent simulations \cite{ref:kob_andersen} showed that for 
present day computer simulations the system falls 
out of equilibrium around $T \approx 0.44$, we run (NVE)-simulations at
temperatures $T=0.15/0.2/0.25/0.3/0.35/0.38/0.4/0.41/0.42$ and $0.43$. To do
so we start with a well equilibrated configuration at $T=0.466$. After
an instantaneous quench to $T=0.15$ we first run a
(NVT)-simulation \cite{ref:Tbathnotes} for $10^5$ MD steps 
to let the system anneal,
and then run the production run with a (NVE) simulation 
for also $10^5$ MD steps. We then increase the
temperature to $T=0.2$ and then again run a (NVT)-simulation followed by a
production run each with $10^5$ MD steps; then increase the temperature
to $T=0.25$ and so forth. In this paper we refer to the so obtained
production runs as ``short runs,'' of which some preliminary results 
have been published elsewhere \cite{ref:proceed_dynhet}. We present here 
mainly results of the so-called ``long runs'' for which we use the 
configurations at the end of the equilibration period of
the short runs but the production runs 
are for $5 \cdot 10^6$ MD steps.
To improve the statistics we run 10 independent configurations
for both long and short runs and for each temperature. 

As it has been demonstrated in earlier work \cite{ref:voll_cool}, the 
structural properties of glasses studied by molecular dynamics
simulation do depend on the preparation history quite distinctly.
Since we study the system out of equilibrium and at finite temperature,
the resulting configurations show some ``aging phenomena''
during the time intervals used for the production of the present data.
However, for those temperatures where even the mean square
displacements of the 5\% fastest particles are still small in
comparison to typical interparticle distances 
over the whole time span of the averaging, this
change of the glass structure due to aging should have a relatively
small effect on our data. As will be seen below, this is the case
for $T \le 0.35$, while for somewhat higher temperatures aging
effects can be expected to become important. The reason for this
is that $\tau$, the typical relaxation time of the system {\em in
equilibrium}, is at $T=0.446$ around $800,000$ time units ($=4\cdot 10^7$
MD~steps)~\cite{ref:gleim98}, thus only a factor 10 longer than the long runs
in the present work. Since in equilibrium the $\alpha-$relaxation time
corresponds to the typical time scale on which a substantial fraction
of the particles have moved (40-70\%), it can be expected that quite
a few particles will show relaxation even on time scales significantly
shorter than $\tau$. Similar relaxation processes are 
also expected  in the out of equilibrium
situation, i.e.  in the glass and thus we do indeed expect aging effects
at temperatures slightly below the (kinetic) glass transition. 
%
We shall comment on this problem that there is some aging
of the glass structure occurring when appropriate.

\section{Mobile and Immobile Particles} \label{sec:mobile_immobile}

\begin{figure}[tbp]
\epsfxsize=3.2in
\epsfbox{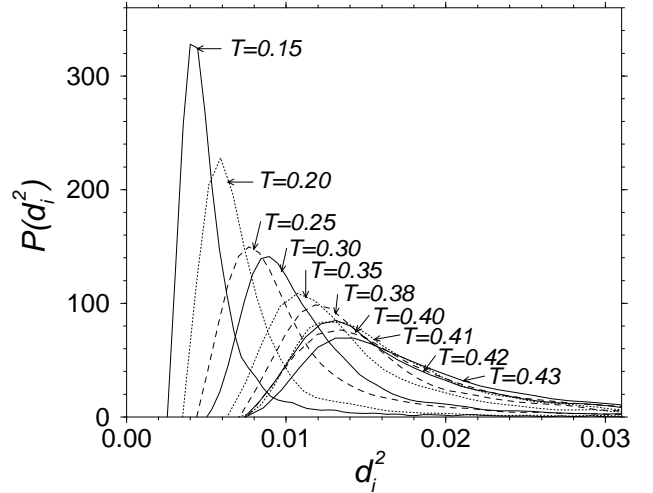}
\caption{The distribution $P(d_i^2)$  for the A
  particles.}
\label{fig:Pofdisq}
\end{figure}

Similar to previous work on dynamic heterogeneities
\cite{ref:kegel,ref:eweeks,ref:schroeder,ref:nistwork}, 
we study the dynamics of the system
by observing the fastest (mobile) and the slowest (immobile)
particles. Since the focus of this work is, however, on the dynamics of
the glass {\em below} the glass transition, our definition of mobile
and immobile particles is different. We have in mind the
picture of a harmonic solid for which the vibrational amplitude 
carries essential information about the local dynamics. We therefore
characterize the mobility of each particle $i$ by 
\begin{equation} \label{eq:disq}
d_i^2 = \overline{ \left |\vec{r}_i - \overline{\vec{r}_i} \right |^2 }
\end{equation}
where the bar denotes an average over a certain time interval. 
We call the 5\% A particles
and separately the 5\% B particles 
with the largest/smallest $d_i^2$ the mobile/immobile particles. 
With ``mobile'' we intend to indicate that these particles are 
relative to all other particles more mobile, they are however in the 
results, presented here, in most cases still bound to their site.
All results presented below are qualitatively the same if the 5\% are
replaced by 10\% particles and are therefore independent of the
specific percentage used. 
The results depend however on the length of the time average,
as will be discussed in detail in section \ref{sec:timescale}.
We use in this paper the non-Gaussian parameter 
\cite{ref:alpha2_1,ref:nistwork,ref:alpha2_2} to determine the length of the 
time average as follows. The non-Gaussian parameter is defined as

\begin{equation} \label{eq:alpha2}
\alpha_2(t) = \frac{3 \, \langle r^4(t) \rangle}
                   {5 \, \langle r^2(t) \rangle^2} \, - \, 1
\end{equation}
where $\langle \cdot \rangle$ corresponds not to the canonical average
since the system is out of equilibrium. Instead we mean by 
$\langle \cdot \rangle$ here and in the following an average over 
particles and initial configurations, i.e. 
\begin{equation}
\langle r^{2n}(t) \rangle = \frac{1}{N} \, \left \langle \sum \limits_i
  \Big | \vec{r}_i(t) - \vec{r}_i(0) \Big |^{2n} \right \rangle
\end{equation}
where the sum goes
over either all  A particles (to obtain $\alpha_\ind{2A}$)
or all  B particles (to obtain $\alpha_\ind{2B}$). This parameter vanishes if
the van Hove correlation function
\begin{equation}
G_s(\vec{r},t) = \frac{1}{N} \, \left \langle \sum \limits_i
   \delta(\vec{r} - [\vec{r}_i(t) - \vec{r}_i(0)]) \right\rangle
\end{equation}
is equal to a Gaussian \cite{ref:alpha=0}  
\begin{equation}  \label{eq:vanHove_gauss}
G_s(r,t) = \left ( \frac{3}{2 \pi \langle r^2(t) \rangle } \right )^{3/2}
             \exp \left ( - 3 r^2 / ( 2 \langle r^2(t) \rangle )
             \right )
\hspace*{2mm}.
\end{equation}
\begin{figure}[tbp]
\epsfxsize=3.2in
\epsfbox{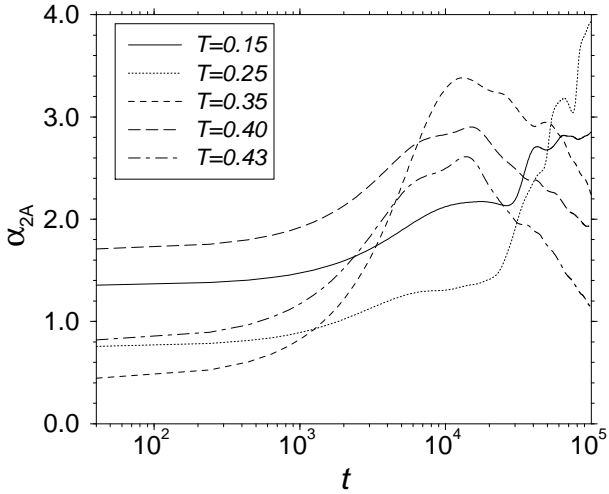}
\caption{Non-Gaussian parameter $\alpha_2(t)$ (see
  Eq.~(\ref{eq:alpha2})) for A particles. For clarity only a subset of 
  all simulated temperatures is shown.
}
\label{fig:alpha2A}
\end{figure}
We expect Eq.~(\ref{eq:vanHove_gauss}), and therefore $\alpha_2=0$, to be
a good approximation for $t \to 0$ (because this corresponds to the
ballistic regime where $r(t) \propto v \cdot t$ which is Maxwell
distributed \cite{ref:alpha2_2}) as well as for $t \to \infty$
(diffusive behavior).  
For intermediate times $\alpha_2(t) \neq 0$ (see Fig.~\ref{fig:alpha2A}). 
Although the non monotonous temperature dependence of $\alpha_2(t)$ 
in Fig.~\ref{fig:alpha2A} 
indicates that our statistics is not very good, we use 
the time $t_\ind{max}$ where $\alpha_2(t)$ reaches its
maximum as an estimate for the characteristic time scale of
mobility. (Note that for $T \leq 0.25$ we use $t_\ind{max}=10^5$.)
In all following results which involve time averages, we use
$t_\ind{max}$, if not otherwise stated, as the time length over which
we average. Note that $t_\ind{max}$ is much larger than the
microscopic oscillation time which is of the order of 1.0. 
We obtain thus for each temperature and for either all A
or all B particles the distribution of $d_i^2$ defined in
Eq.(\ref{eq:disq}). Fig.~\ref{fig:Pofdisq} shows for the A
particles that with increasing
temperature the distribution shifts to the right and develops a longer tail. 
(Similar results are obtained for the B particles.)
The curves are zero for $d_i^2 < 0.0025$, which reflects the fact that all
particles are oscillating somewhat. The tail of $P(d_i^2)$ extends for
high temperatures to values of $d_i^2$ that are twice as large as 
the $d_i^2$ at the peak position, which
shows that the dynamics is rather heterogeneous.
We also mention that the  $P(d_i^2$) for different $T$ 
can be collapsed onto a single
curve by rescaling the distribution to $P(d_i^2/\langle d_i^2
\rangle)$ (see \cite{ref:kamal_Pdisq}). 

\begin{figure}[tbp]
\epsfxsize=3.2in
\epsfbox{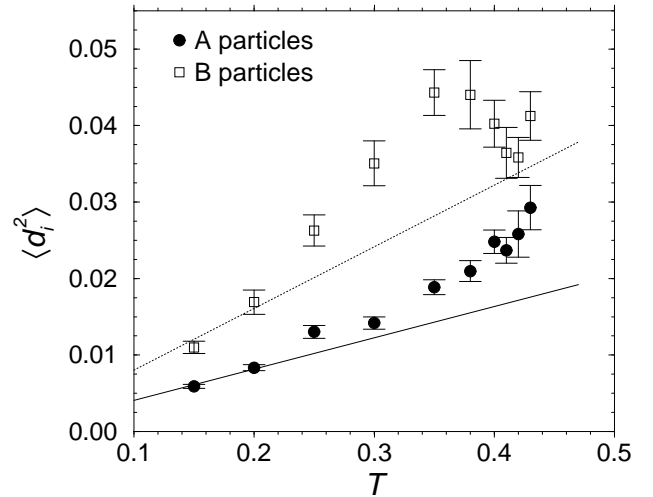}
\caption{Temperature dependence of $\langle d_i^2 \rangle$ 
for the A and B particles. The straight lines are fit to the data 
in the harmonic regime.}
\label{fig:disqav}
\end{figure}
\begin{figure}[tbp]
\epsfxsize=3.2in
\epsfbox{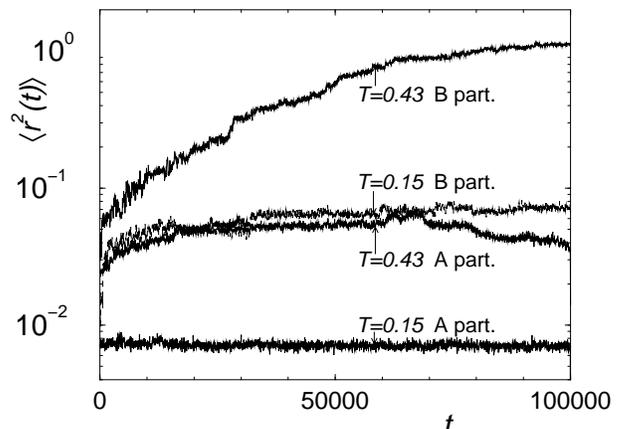}
\caption{$\langle r^2(t) \rangle$ for the 
 slowest 5\% A particles and B particles.
}
\label{fig:msd_slow}
\end{figure}
\begin{figure}[tbp]
\epsfxsize=3.2in
\epsfbox{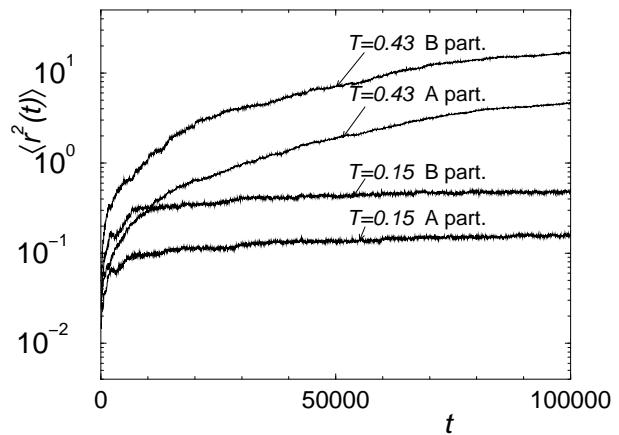}
\caption{$\langle r^2(t) \rangle$ for the 
 fastest 5\% A particles and B particles.
}
\label{fig:msd_fast}
\end{figure}

Fig.~\ref{fig:disqav} summarizes the average 
values of $d_i^2$ for the A and B particles
$\langle d_i^2 \rangle$ as a function of temperature. 
For a harmonic system one would expect a linear dependence of 
$\langle d_i^2 \rangle$ through the origin and 
over the whole range of temperatures. 
Since the B particles are smaller, they have a larger amplitude 
of oscillation than the A particles. 
For very small temperatures $\langle d_i^2 \rangle$ increases linearly and 
deviates for A and B particles from a line for larger temperatures.
The decrease of 
$\langle d_i^2 \rangle$ of B particles for increasing temperature 
at high temperatures is 
due to our time average with $t_\ind{max}$. If one averages instead over the
complete long simulation run, $\langle d_i^2 \rangle$ increases 
monotonically and even more than linearly.
The discrepancies from a straight line through the origin 
for A and B particles at temperatures $T \gtrsim 0.25$ show that 
anharmonic effects become important already at small temperatures.
As suggested in \cite{ref:angell_shao}, this onset of anharmonicity may
be related to the calorimetric glass transition.
\begin{figure}[tbp]
\epsfxsize=3.2in
\epsfbox{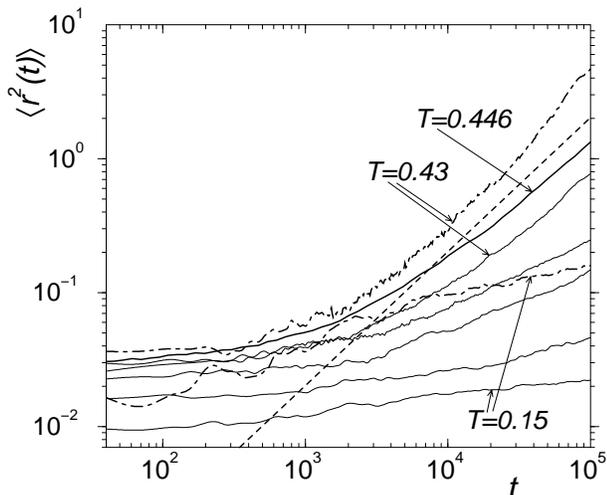}
\caption{$\langle r^2(t) \rangle$ for 
all A particles at the temperatures 0.15/0.25/0.35/0.4/0.43 (solid lines) and 
for comparison the equilibrium data  at $T=0.446$ (bold solid line). 
Included are also the $\langle r^2(t) \rangle$ 
for the fastest 5\% A particles (bold dot-dashed lines)
and a bold dashed line of slope 1. 
}
\label{fig:msd_allA}
\end{figure}
The log-log plot of the mean square displacement 

\begin{equation}
\langle r^2(t) \rangle = \frac{1}{N_\ind{A}} \, 
  \left \langle \sum \limits_{i=1}^{N_{\mbox{\tiny A}}} 
   \left | \vec{r}_i(t) - \vec{r}_i(0) \right |^2 \right \rangle
\hspace*{2mm},
\def \ind#1{{\mbox{\scriptsize {#1}}}}
\end{equation}
where we average over all A particles (see Fig.~\ref{fig:msd_allA}),
shows that the slopes at large times are smaller 
than one. 
Therefore most particles never reach the
diffusive region ($m=1$) and are trapped in their cages during the whole
simulation run at least for $T \leq 0.35$.
We find the same for
the B particles where $m  \lesssim 0.93$. 
If we average over only the 5\% particles with the
largest $\langle r^2(t_\ind{end}=10^5) \rangle$, the late time slopes are
at $T=0.43$ $m \approx 1.4$ 
for the A and $m \approx 1.2 $ 
for the B particles. 
This transient behavior of $m > 1$ might be due to jump processes.
Jumps are clearly visible for the B-particles in Fig.~\ref{fig:msd_slow}, 
which shows the 5\% particles with the smallest 
$\langle r^2(t_\ind{end}) \rangle $.
At low temperatures the slowest A particles are trapped at their site 
as can be seen in Fig.~\ref{fig:msd_slow},
since $\langle r^2(t)\rangle < 10^{-2}$ over the whole simulation run.
Note also that the slowest B particles are faster than the average A
particles. Fast B particles at $T=0.43$ are reaching values of even 
$\langle r^2(t_\ind{end}) \rangle > 10$ (see Fig.~\ref{fig:msd_fast}).
%

\section{Spatial Distribution} \label{sec:spatialdist}

\begin{figure}[tbp]
\epsfxsize=3.2in
\epsfbox{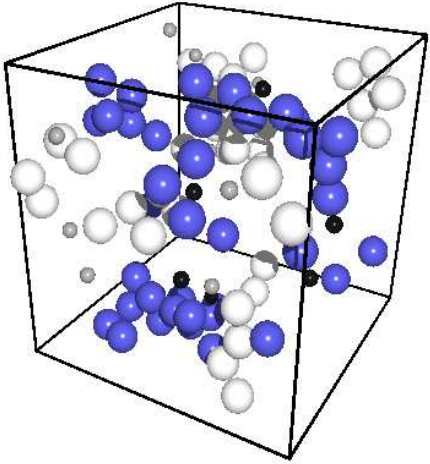}
\caption{Snapshot of the mobile A (white large spheres) and 
B particles (light grey small spheres)  and the immobile A (dark grey large 
spheres) and B particles (black small spheres)
at $T=0.15$ and at the beginning of the production run.
The radii were chosen for clarity and do not reflect the parameters of
the potential.} 
\label{fig:snap_150n_01_t0}
\end{figure}

\begin{figure}[tbp]
\epsfxsize=3.2in
\epsfbox{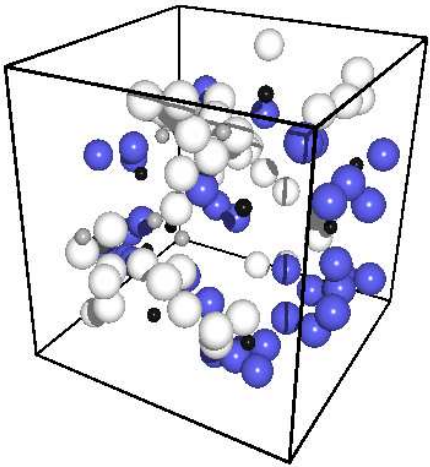}
\caption{Same as Fig.~\ref{fig:snap_150n_01_t0} but for $T=0.43$}
\label{fig:snap_430n_01_t0}
\end{figure}

With the definition of mobile and immobile particles as given in the
last section, we study now how they are spatially 
distributed. Fig.~\ref{fig:snap_150n_01_t0} and \ref{fig:snap_430n_01_t0}
show the spatial distribution of the mobile (light spheres)
and immobile (dark spheres) 
at temperatures $T=0.15$ and $T=0.43$, and at the beginning of the 
time interval for which their mobilities are determined.  For clarity all
other 900 particles are not shown. 
In these snapshots, and similarly for all other temperatures and times,
the particles are clearly distributed in a
heterogeneous way. We therefore find dynamic heterogeneity for
all investigated states in the glass phase.
The number of particles in the largest cluster \cite{ref:clusterdef} 
is at all investigated 
temperatures for mobile particles about 30 particles and for 
immobile particles about 22 particles.
These clusters are smaller than the clusters of Weeks {\it et al.} 
\cite{ref:eweeks}. The likely reason for this discrepancy is that we study a 
smaller system and with less good statistics.

\begin{figure}[tbp]
\epsfxsize=3.2in
\epsfbox{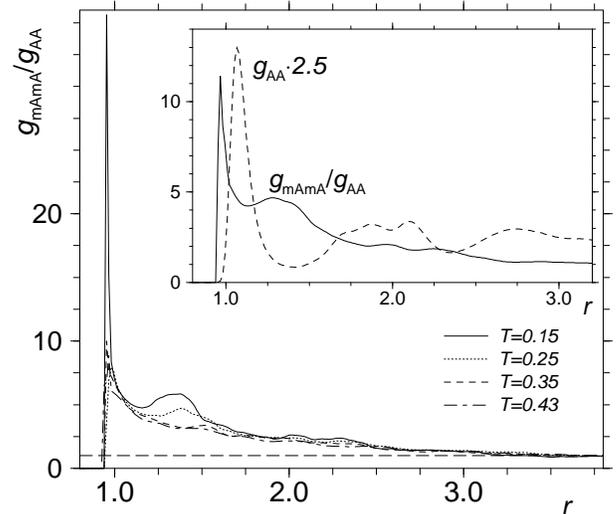}
\caption{$g_\ind{mAmA}/g_\ind{AA} (r)$ 
  at different temperatures. The horizontal dashed line 
  at $g_\ind{mAmA}/g_\ind{AA}=1$ is for the guidance of the eye.
 The inset is a comparison of $g_\ind{mAmA}/g_\ind{AA}$ (solid line) 
  and $g_\ind{AA}(r)\cdot 2.5$ (dashed line) at $T=0.2$.}
\label{fig:gmAmA_tcut}
\end{figure}

To quantify the spatial heterogeneity we plot similar to \cite{ref:nistwork}
the ratio $g_\ind{mAmA}/g_\ind{AA}$  between the radial pair distribution 
\cite{ref:allen+tildesley,ref:hansen+mcdonald}
of solely mobile particles  and that of all particles
(Fig.~\ref{fig:gmAmA_tcut}) with 
\begin{equation}
g(r) = \frac{V}{N^2} \, \langle \sum_i \sum_{j \ne i} \,
   \delta \left (\vec{r}-[\vec{r}_i-\vec{r}_j] \right ) \rangle
\hspace*{2mm}.
\end{equation}
%
In the case of randomly selected 5\% particles from all A particles,
this ratio would be one.
We find however that
this ratio is not a constant with respect to $r$
(and similarly for the corresponding ratios of AB and BB), 
which confirms the dynamic heterogeneity. Since $g_\ind{mAmA}/g_\ind{AA}>1.0$
for distances $r \lesssim 3.2$, 
mobile particles tend to be near each
other. We can conclude from the position of the first peak of
$g_\ind{mAmA}/g_\ind{AA}$ (see inset of Fig.~\ref{fig:gmAmA_tcut}) 
that separation distances
which are for average particles very unlikely, as at the left wing of
$g_\ind{AA}$, occur for mobile particles more often. 

\begin{figure}[tbp]
\epsfxsize=3.2in  
\epsfbox{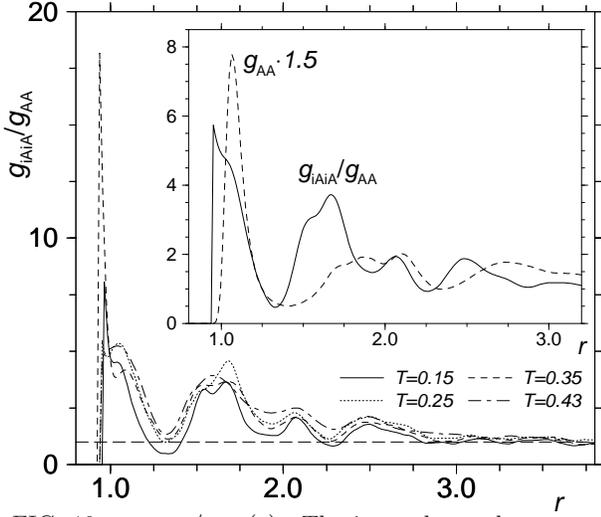}
\caption{
$g_\ind{iAiA}/g_\ind{AA} (r)$. The inset shows the comparison with
$g_\ind{AA}$.  The symbols are the same as in  Fig.~\ref{fig:gmAmA_tcut}.}
\label{fig:glAlA_tcut}
\end{figure}
We can draw similar conclusions for the immobile particles. 
Fig.~\ref{fig:glAlA_tcut}
shows that the ratio of the radial pair distribution of immobile
particles to that of all particles, $g_\ind{iAiA}/g_\ind{AA}$, is also
larger than one for small distances. The inset, which includes
$g_\ind{AA}$ for comparison, reflects that also for immobile particles
very small separation distances are more likely.

\section{Surrounding}\label{sec:surrounding}

In the last section we found that the mobile/immobile particles 
form clusters and that the typical distances between the particles 
are different from those in the bulk.  We now address
the question of the reason for mobility by comparing the
surrounding of mobile/immobile particles to the one of
average particles.

\subsection{Coordination Numbers}

\begin{figure}[tbp]
\epsfxsize=3.2in
\epsfbox{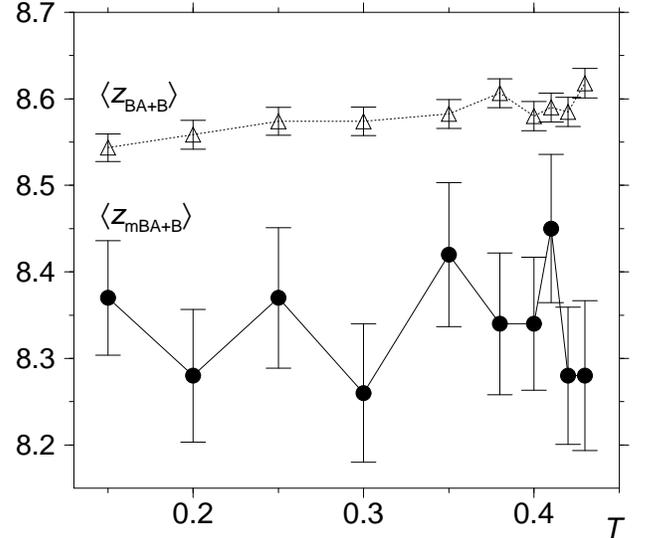}
\caption{Number of neighbors (counting both 
A and B particles) of a mobile B particle, $\langle z_\ind{mBA+B}\rangle$, 
in comparison to the number of neighbors of an average B particle, 
$\langle z_\ind{BA+B} \rangle$.}
\label{fig:zmBA+B_tcut}
\end{figure}

\begin{figure}[tbp]
\epsfxsize=3.2in
\epsfbox{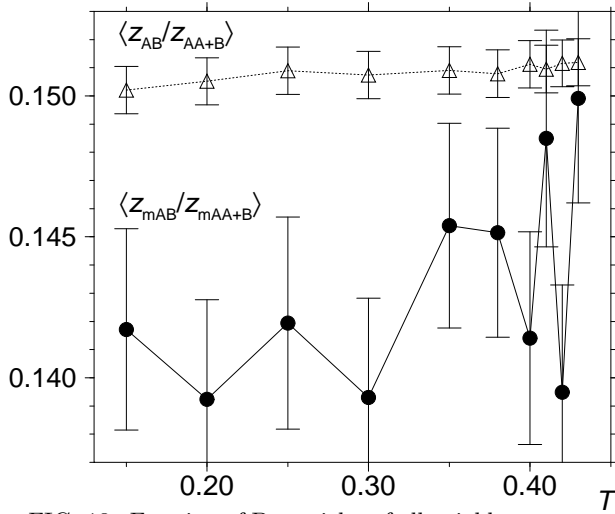}
\caption{Fraction of B particles of 
all neighbors surrounding a mobile A particle (filled circle) and an
average A particle (open triangle).}
\label{fig:zmABdzmAA+B_tcut}
\end{figure}

To probe the immediate neighborhood of the mobile and immobile
particles we count their number of nearest neighbors (coordination
number $z$) where a particle $j$ is defined to be a neighbor of
particle $i$ if their distance $|\vec{r}_{ij}|=|\vec{r}_i-\vec{r}_j|$ is
smaller than the position of the first minimum $r_\ind{min}$ 
of the corresponding (average) radial pair distribution function
($r_\ind{min}=1.4$ for AA, $1.2$ for AB and $1.07$ for BB, independent 
of temperatures). 
\begin{figure}[tbp]
\epsfxsize=3.2in
\epsfbox{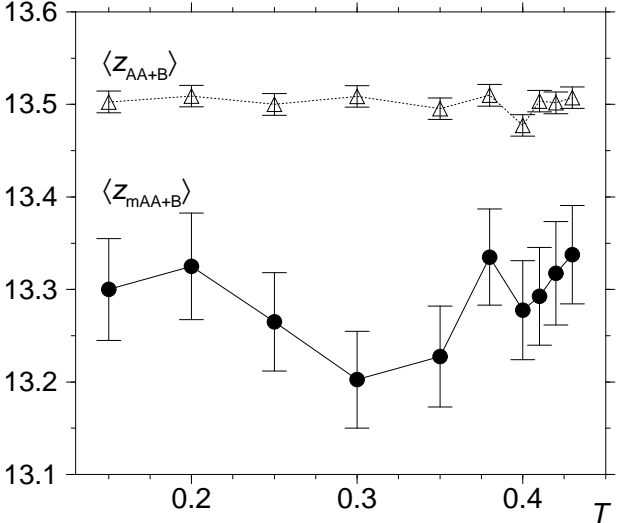}
\caption{Number of neighbors (counting both 
A and B particles) of a mobile A particle, $\langle z_\ind{mAA+B}\rangle$, 
in comparison to the number of neighbors of an average A particle, 
$\langle z_\ind{AA+B} \rangle$.}
\label{fig:zmAA+B_tcut}
\end{figure}
Fig.~\ref{fig:zmAA+B_tcut} shows that a mobile A
particle is on average surrounded by fewer particles, $\langle
z_\ind{mAA+B} \rangle$, than an average A particle, $\langle
z_\ind{AA+B} \rangle$. This suggests that one of the reasons for 
a particle to be mobile is that it is on average caged by fewer
particles. The same is true for mobile B particles 
(see Fig.~\ref{fig:zmBA+B_tcut}). 
However, we cannot make the stronger statement that any A particle with 
$\langle z_\ind{mAA+B} \rangle $ less than, say, $13.4$ is mobile, because the 
distribution of coordination numbers is quite broad with 
a standard deviation of $\sigma \approx 2.0$.
  Mobile A particles are furthermore surrounded by a lower than average
percentage of B neighbors (see Fig.~\ref{fig:zmABdzmAA+B_tcut}), 
because the latter
trap A particles both energetically ($\epsilon_\ind{AB} >
\epsilon_\ind{AA}$) as well as geometrically ($\sigma_\ind{AB} <
\sigma_\ind{AA}$). 

\begin{figure}[tbp]
\epsfxsize=3.2in
\epsfbox{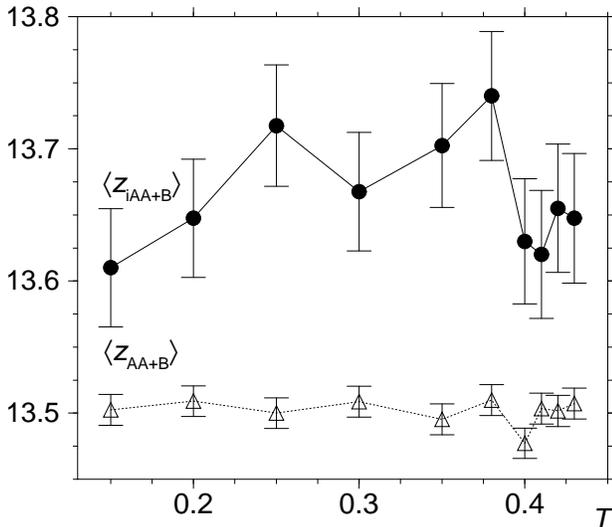}
\caption{Total number of neighbors of an 
immobile A particle (filled circle) and an average A particle (open triangle).}
\label{fig:zlAA+B_tcut}
\end{figure}

\begin{figure}[tbp]
\epsfxsize=3.2in
\epsfbox{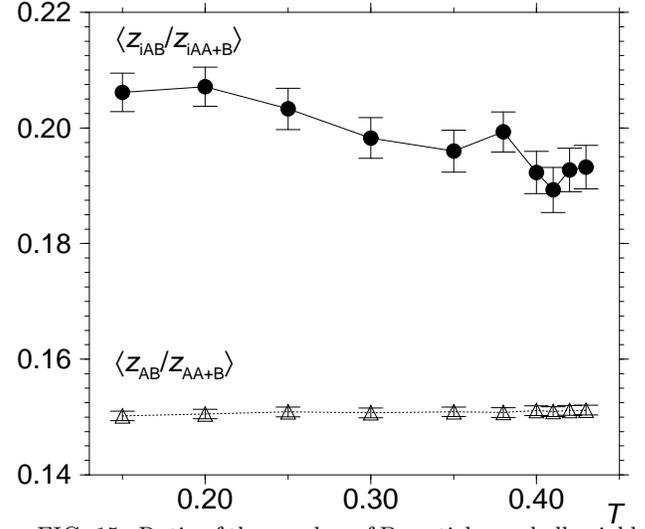}
\caption{Ratio of the number of B particles
  and all neighbors of an immobile A particle (filled circle) 
  in comparison with the ratio of the number of B particles and all
  neighbors of any A particle (open triangle).}
\label{fig:zlABdzlAA+B_tcut}
\end{figure}

\begin{figure}[tbp]
\epsfxsize=3.2in
\epsfbox{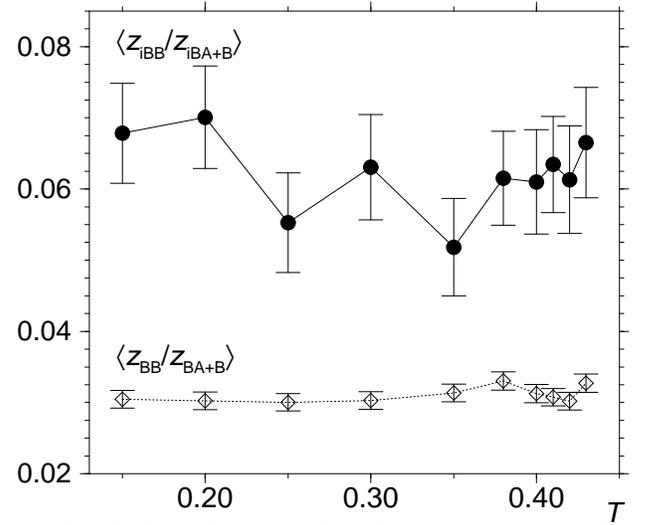}
\caption{Same figure as Fig.~\ref{fig:zlABdzlAA+B_tcut}
 but now for the neighbors of an immobile B particle.}
\label{fig:zlBBdzlBA+B_tcut}
\end{figure}

Similarly immobile particles have the property to
have more neighbors than average particles
(see Fig.~\ref{fig:zlAA+B_tcut})
and of a higher percentage of B particles than usual 
(see Fig.~\ref{fig:zlABdzlAA+B_tcut} and \ref{fig:zlBBdzlBA+B_tcut}). 
Notice that the latter is true both for A and B particles (see
Fig.~\ref{fig:zlABdzlAA+B_tcut} and \ref{fig:zlBBdzlBA+B_tcut}) 
due to the tighter packing with the smaller B
particles. 

We also find that the percentage of 
mobile/immobile neighbors of a mobile/immobile particle
is significantly larger than 5\%, i.e., 
$\langle z_\ind{mAmA+mB} \rangle > 0.05 \cdot 
\langle z_\ind{mAA+B} \rangle$ and $\langle z_\ind{iAiA+iB} \rangle > 0.05 \cdot
\langle z_\ind{iAA+B} \rangle$,
which reflects once more the spatial heterogeneity discussed in 
Fig.~\ref{fig:gmAmA_tcut} and \ref{fig:glAlA_tcut}.

\subsection{Radial Pair Distribution Functions}

\begin{figure}[tbp]
\epsfxsize=3.2in
\epsfbox{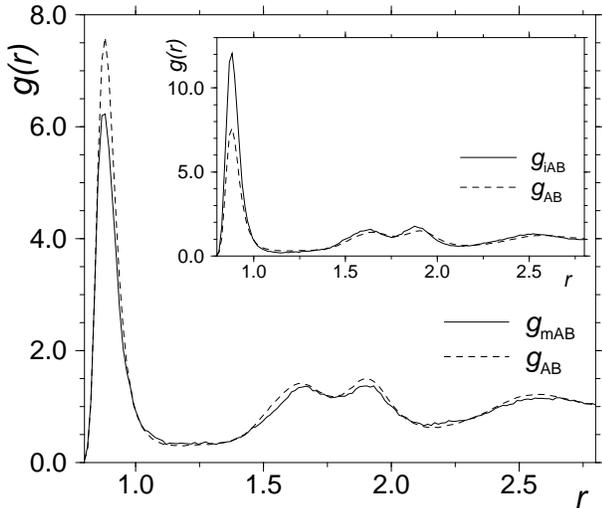}
\caption{Radial pair distribution function of a mobile 
A particle and any B particle (solid line) 
and for comparison the corresponding 
radial distribution function of any A  and B particles (dashed line).
The inset shows the similar $g(r)$ for the immobile A particle with any 
B particle. Both graphs are for $T=0.15$.}
\label{fig:gmAB+lAB}
\end{figure}

Next we use the radial pair distribution function to study the environment
of the mobile and immobile particles beyond the nearest neighbor shell.
Fig.~\ref{fig:gmAB+lAB} 
shows the radial pair distribution function of a mobile A particle
with any B particle, $g_\ind{mAB}$, in comparison with $g(r)$ of 
any A and B particles, $g_\ind{AB}$, at $T=0.15$. 
We find that $g_\ind{mAB}$ has smaller maxima and broader peaks than 
$g_\ind{AB}$ which corresponds, specifically for the first neighbor shell,
to an effectively wider cage around the mobile particles
\cite{ref:footnote_laird}. 
The wider cage allows larger distances and thus larger $d_i^2$ which
corresponds by definition to mobile particles.
We see the same effect for $g_\ind{mAA}$ 
(for $g_\ind{mBB}$ the statistics is not sufficient) and for all other
temperatures. 

Immobile particles, in contrast, are surrounded by an effectively
narrower cage, as can be concluded from the more pronounced peak 
of the first maximum (see inset of Fig.~\ref{fig:gmAB+lAB}). 
This is observed for all
temperatures and all radial pair distribution functions characterizing
the surrounding of an immobile particle.
Notice that the change of the neighborhood is larger around an immobile 
particle than around a mobile particle, as the comparison of 
Fig.~\ref{fig:zmABdzmAA+B_tcut} and Fig.~\ref{fig:zlABdzlAA+B_tcut}
and the comparison of Fig.~\ref{fig:gmAB+lAB} and its inset show. 
This is probably due to our definition of mobility: since the
distribution of $d_i^2$ (see Fig.~\ref{fig:Pofdisq}) is very asymmetric,
5\% particles with the smallest $d_i^2$ cover a much smaller range of
$d_i^2$ than 5\% particles with the largest $d_i^2$. Immobile particles
are thus more distinct than mobile particles.

\section{Time Scale of Mobile and Immobile Particles} \label{sec:timescale}

In this section we get back to Eq.~(\ref{eq:disq}), which is essential
for the definition of mobile and immobile particles. We vary the time
length over which we average. Specifically, we average over the simulation 
time for the long and short runs (see Sec.~\ref{sec:simulation}), 
rather than using $\alpha_2$ 
to determine a temperature-dependent time for calculating the average.
We now investigate the influence of this averaging time on the results
presented in the previous three sections.

\begin{figure}[tbp]
\epsfxsize=3.2in
\epsfbox{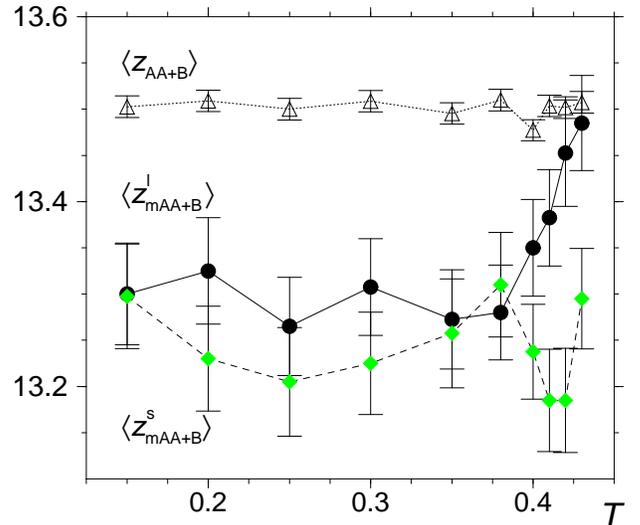}
\caption{Total number of neighbors of any 
A particle $\langle z_\ind{AA+B}\rangle$ (open triangle) and of a 
mobile particle defined for the longer simulation run 
$\langle z_\ind{mAA+B}^\ind{l} \rangle$ (dark filled circle) and the shorter 
simulation run $\langle z_\ind{mAA+B}^\ind{s} \rangle$ (grey diamond).}
\label{fig:zmAA+Bls}
\end{figure}
Fig.~\ref{fig:zmAA+Bls} shows a comparison of the 
total number of nearest neighbors of a
mobile  A particle for the short runs, $\langle z_\ind{mAA+B}^\ind{s}
\rangle$, and the long runs, $\langle z_\ind{mAA+B}^\ind{l} \rangle$. All
coordination numbers have been averaged over the 10 independent initial
configurations, and no data from later times are included. 
The difference in $\langle
z_\ind{mAA+B}^\ind{s} \rangle$ and $\langle z_\ind{mAA+B}^\ind{l} \rangle$ is
solely due to the different definition of mobile particles. As before
(see Fig.~\ref{fig:zmAA+B_tcut}) 
we find that the mobile particles are surrounded by fewer
particles. For the long runs, however, this effect vanishes at higher
temperatures. The likely reason for this decreasing difference is that
we used in Eq.~(\ref{eq:disq}) a time average over
the entire long simulation run, while a fast particle
might be fast only over some fraction of the simulation
run. While the particle is fast, its environment is different than that
of a regular particle. However, when the time average includes
also times when the particle is not fast, then we dilute the average with
environments that are not
special. This mixing with the environment of average particles happens
more readily at temperatures $T \gtrsim 0.4$ when we approach the glass
transition  because the typical time scale over which a particle is
fast is shorter than at lower temperatures. 

\begin{figure}[tbp]
\epsfxsize=3.2in
\epsfbox{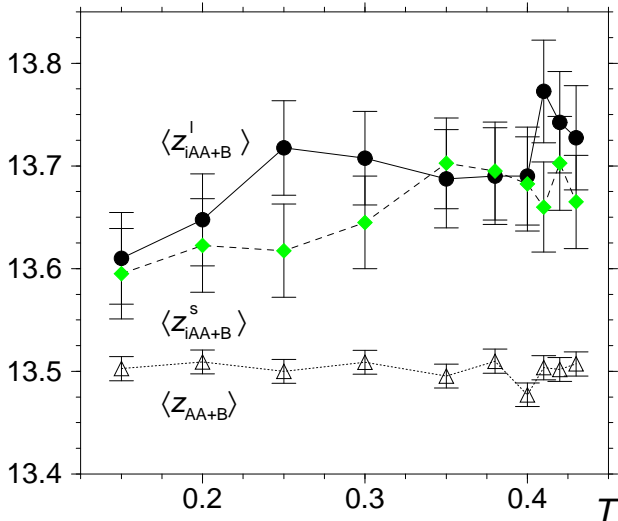}
\caption{Total number of neighbors of an average 
A particle $\langle z_\ind{AA+B}\rangle$ (open triangle) 
and of an immobile particle defined for 
the longer simulation run $\langle z_\ind{iAA+B}^\ind{l} \rangle$
(dark filled circle) and the shorter simulation run 
$\langle z_\ind{iAA+B}^\ind{s} \rangle$ (grey diamond).}
\label{fig:zlAA+Bls}
\end{figure}
Other coordination numbers show the same behavior with the
exception of immobile A particles (see
Fig.~\ref{fig:zlAA+Bls}). The latter
distinguish themselves from the average particle 
even at high temperature and thus
are immobile over the whole simulation run (consistent
with Fig.~\ref{fig:msd_slow}). 

\begin{figure}[tbp]
\epsfxsize=3.2in
\epsfbox{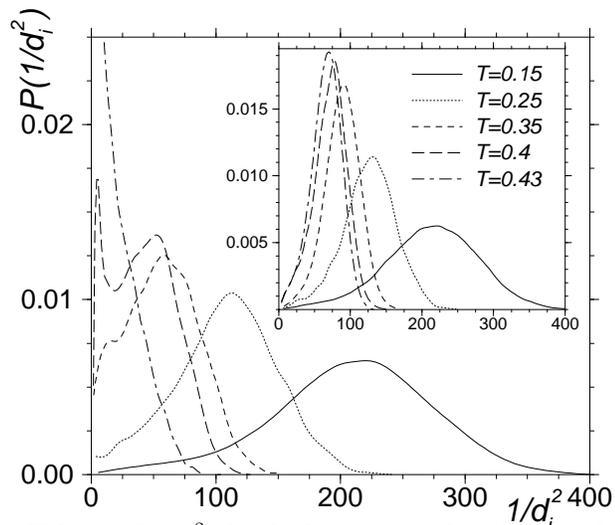}
\caption{$P(1/d_i^2)$ for the long runs and
in the inset for the short runs at various temperatures.}
\label{fig:inv_dist}
\end{figure}
Another quantity which is also strongly dependent on the time length
of the run is the distribution of $1/d_i^2$ 
(see Fig.~\ref{fig:inv_dist} for long runs
and its inset for short runs).
In the case of long runs a double peak structure develops for increasing
temperature. We tentatively associate the particles in the peak with
large $1/d_i^2$ with localized particles and those in the peak with small
$1/d_i^2$ with mobile particles. Further work is required to check this
hypothesis.

The peak at small $1/d_i^2$ 
starts to dominate with increasing temperatures, which
reflects that at high temperatures most particles have become fast at
some time during the simulation run.

As the figures of Secs.~\ref{sec:spatialdist} and
\ref{sec:surrounding} demonstrate, these effects of averaging over fast
and regular particles can be avoided via appropriate choice of averaging time
$t_\ind{max}$ which we chose to be 
when $\alpha_2(t)$ reaches its maximum. Therefore
$t_\ind{max}$ gives us a rough estimate about the lifetime of fast
particles.

\section{Summary and Outlook} \label{sec:summary}

We investigate the mobile and immobile particles of a glass. Note that
our definition of mobile and immobile is different than the
definition given in \cite{ref:nistwork} since we have the picture
of a solid in mind (similar to G. Johnson {\it et al.} who study solid-like
particle clusters \cite{ref:gould}).

We find {\em below} the glass transition temperature a clear dynamic
heterogeneity which has previously been seen in simulations {\em
above} $T_g$ 
and experiments below $T_g$.
To address the question why certain particles
are more/less mobile than others, we study their
surrounding. As one might have expected, the mobile/immobile particles
are surrounded by fewer/more neighbors, forming a cage which is
effectively wider/narrower than the one of a regular particle. In
addition mobile/immobile particles are trapped by fewer/more B
particles, which are smaller than the A particles and therefore allow
closer packing \cite{ref:footnote_epsAB}. We expect that, similarly, a
surrounding specific to the mobility of the central particle might be
found in the future in experiments. Both the dynamic heterogeneity
as well as the particular surrounding of mobile and immobile particles
are consistent with collective behavior as it has been found above the
glass transition temperature $T_g$. 
A more detailed analysis below $T_g$ is left for future work. 

The characteristics of mobility show a time dependence which is well
estimated with the time $t_\ind{max}$ of the maximum of $\alpha_2$.
We conclude that mobile particles are ``fast'' only for a certain time window of
the simulation run whereas the immobile A particles seem to stay mostly
immobile over the range of our simulation runs.
This raises the question of a more precise criterion for the
time scale of fast and slow processes, which we leave for future
work.

\section{Acknowledgments}

KVL gratefully acknowledges financial support from the SFB 262.


\begin{references}
\bibitem{ref:Tbathnotes} To keep the temperature constant, every 50
  time steps the velocities of all the particles were replaced by new
  velocities which were drawn from the Boltzmann distribution for the 
  corresponding temperature.
\bibitem{ref:alpha=0} Note that the integration for $\langle r^{2n} \rangle$ 
   includes the factor $4 \pi r^2$ due to the 
   three dimensional integration with $G_s(\vec{r},t)$.
\bibitem{ref:clusterdef} Particles are defined to be members of the same
  cluster when they are connected with nearest neighbor distances (see
  Sec.~\ref{sec:surrounding}).
\bibitem{ref:footnote_laird} A similar effect has been found in the
  case of a soft sphere glass at low temperatures for the particles
  participating in a soft mode (see \cite{ref:laird}).
\bibitem{ref:footnote_epsAB} Specific to our potential with 
   $\epsilon_\ind{AB}>\epsilon_\ind{AA}$ are mobile/immobile A particles also
  energetically less/more bound when surrounded by fewer/more B
  particles.
\end{references}
\end{document}